\documentclass[submission,copyright,creativecommons]{eptcs}
\providecommand{\event}{FAVO 2011}
\usepackage{graphicx}
\usepackage{hyperref}
\usepackage{enumitem}
\usepackage{blindtext}

\hypersetup{pdfborder=0 0 0}

\title{Multi-model-based Access Control in Construction Projects}

\author{Frank Hilbert
\institute{TU Dresden}
\institute{Institute of Construction Informatics\\
Dresden University of Technology\\
Dresden, Germany}
\email{frank.hilbert@tu-dresden.de}
\and
Raimar J. Scherer
\institute{TU Dresden}
\institute{Institute of Construction Informatics\\
Dresden University of Technology\\
Dresden, Germany}
\email{raimar.scherer@tu-dresden.de}
\and
Larissa Araujo
\institute{USP}
\institute{Engineering School of Sao Carlos\\
University of Sao Paulo\\
Sao Paulo, Brazil}
\email{larissaedaraujo@sc.usp.br}
}

\def\titlerunning{Multi-model-based Access Control in Construction Projects}
\def\authorrunning{F. Hilbert, R. J. Scherer, L. Araujo}
\begin{document}
\maketitle

\begin{abstract}
During the execution of large scale construction projects performed by Virtual Organizations (VO), relatively complex technical models have to be exchanged between the VO members. For linking the trade and transfer of these models, a so-called multi-model container format was developed. Considering the different skills and tasks of the involved partners, it is not necessary for them to know all the models in every technical detailing. Furthermore, the model size can lead to a delay in communication. In this paper an approach is presented for defining model cut-outs according to the current project context. Dynamic dependencies to the project context as well as static dependencies on the organizational structure are mapped in a context-sensitive rule. As a result, an approach for dynamic filtering of multi-models is obtained which ensures, together with a filtering service, that the involved VO members get a simplified view of complex multi-models as well as sufficient permissions depending on their tasks.
\end{abstract}

\section{Introduction}
For the processing of large and complex construction projects, single independent organizations work together as Virtual Organizations (VO) [1], combining their core competencies [2, 3]. To manage the complexity in construction in which the costs incurred within a project vary, both physical and functional information split into various domain-specific application models (e.g. building model, cost model, time model, etc.) on the basis of which the work is running, typically discipline-specific [4]. The use of Building Information Modelling (BIM) is becoming increasingly important for collaborative processing [5]. The elements of the various specialized models are implicitly linked together (e.g. the component inside wall “xy” from the model building with the process “concrete first floor” from the process model). To transfer application models together with their links, in the German Project MEFISTO
\footnote{http://www.mefisto-bau.de}
and in the international Project HESMOS
\footnote{http://hesmos.eu}
, a multi-model container format (MMC) was developed which summarizes application models with their dependencies [6]. When using this MMC approach, however, the following aspects must be considered:

\begin{itemize}
\item (A1) The MMC, based on the number and granularity of the application models, becomes relatively large. Because of distributed processing in Virtual Organizations, these multi-models are shared very often and the size of the model can lead to the obstruction of communication (e.g. the size of the IFC building model of a simple high-rise building is about 40 MB).

\item(A2) For the processing of individual project tasks only a relatively small part of the subject model is usually interesting. According to their current project responsibilities the agents are only interested, in most cases, in the specific aspects (e.g. building services, building spatial structures, structural analysis, etc.).

\item (A3) It is not necessary, and for reasons of data protection, may also not be desirable that all partners involved must know all the technical details of the project model.
\end{itemize}

From these considerations, it is useful to facilitate the working with multi-models, the transmitting of model cut-outs or the generating of model views. For this purpose, a central filter and mapping service are currently being developed at the TU Dresden, which generate partial application models while maintaining the link structure between the application models [7]. This filtering service is used within this approach to enforce the definition of model views and covers the aspect of access and privacy when accessing multi models within virtual organizations. In this manner, an initiated approach, as to the integration of these filter services in a context-sensitive fine-grained access control, is described. The next section briefly describes the structure of the multi-model used. On this basis, in Chapter 3, different multi-model filter designs are considered; first the uses of multi-model templates for the generation of model views, and second, the filtering of elementary models for creating cut-outs. Then, in chapter 4, the idea of an extended context-sensitive role model for virtual organizations is pursued, and in chapter 5, a proposed methodology for using this model for the filtered access to multi-models follows. The last chapter outlines practical scenarios and gives an outlook into further development.

\section{Basic Concepts}
\subsection{The multi-model container format}
In order to distribute project information together from the various semantically and structurally inhomogeneous domain expert models and, on the other side, externalize the implicit relationship between the model elements, a generic link model was developed, which will be shown together with the associated application models within a multi-model. Then the link model in the form of an XML document is stored, the format of the application models is not specified. The multi-model was changed in a so-called multi-model container (MMC) which is within the multi-model container and separated in each model, described by metadata (see Figure 1). These multi-model containers are read into the specialized applications of the VO partners, modified and shared with other partners. More information about the structure and the use of MMC is given by Fuchs in [6].

\begin{figure}[htb]
\centering
\includegraphics[scale=0.7]{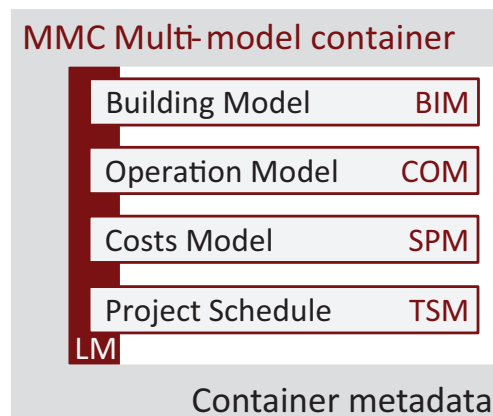}
\centering
\caption{Example Structure of the Multi-Model-Container}
\end{figure}

\subsection{Filter Concepts for Multi-Models}
In the filtering of multi-models it is necessary to distinguish between the pure selection of application models as they are defined for model-views (domain-views) with any setting of the desired granularity (see Figure 2) and of the filter model to generate model cut-outs (see Figure 3).

\begin{figure}[htb]
\begin{minipage}[htb]{0.475\textwidth}
\centering
\includegraphics[scale=0.7]{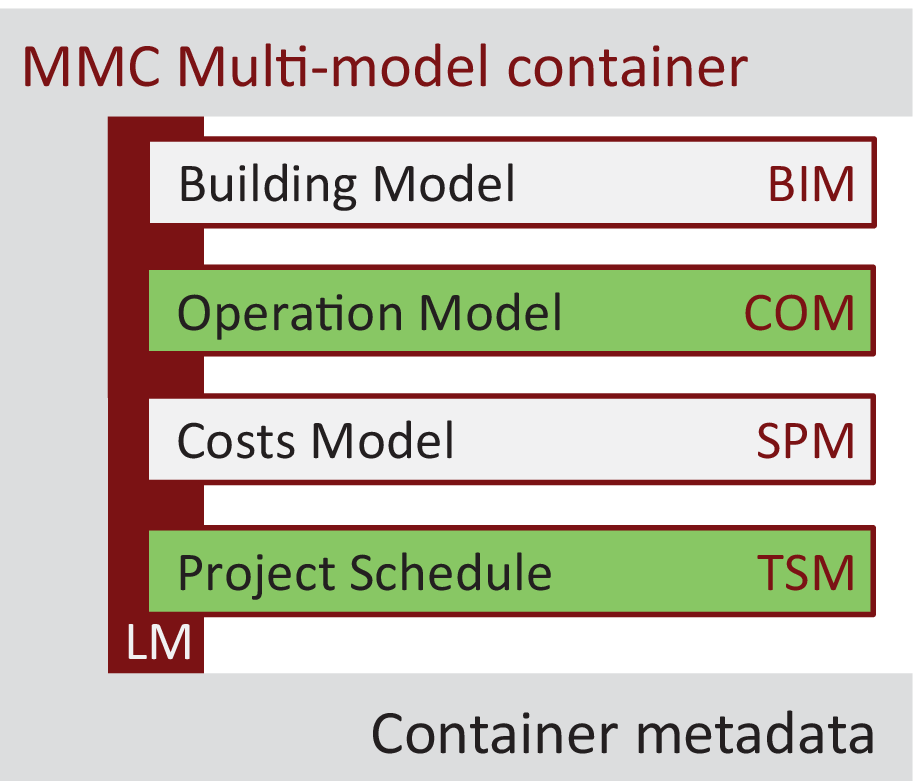}
   \caption{Example of a Multi-Model-View}
    \label{Bild_A}
\end{minipage}
\hfill
\begin{minipage}[htb]{0.475\textwidth}
\centering
\includegraphics[scale=0.7]{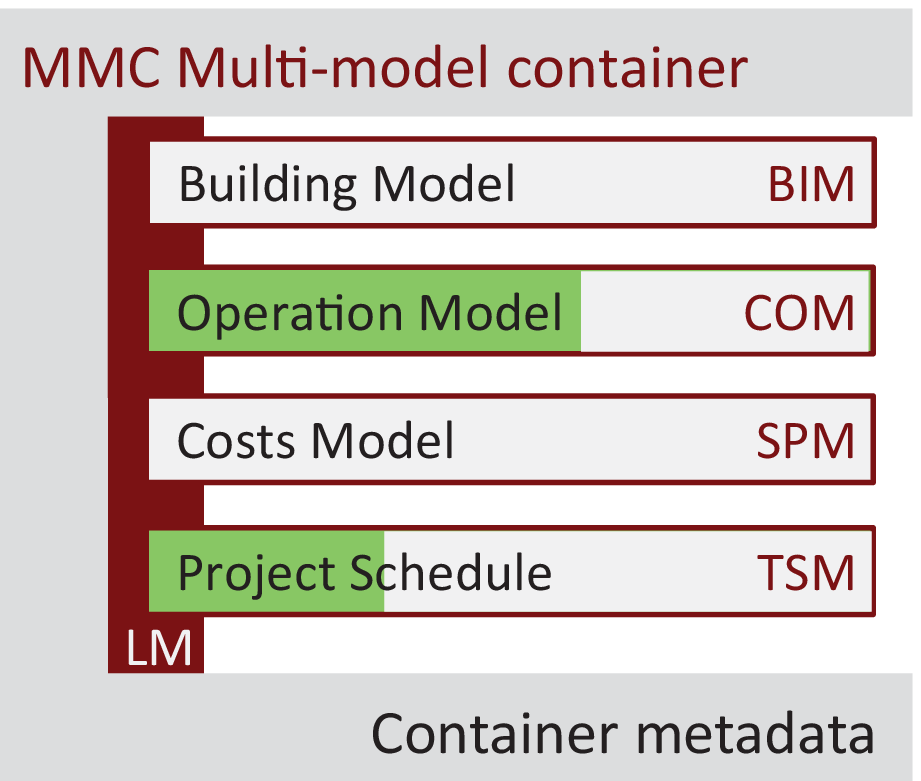}
\caption{Example of a Multi-Model-Cutout}
    \label{Bild_B}
\end{minipage}
\end{figure}

For handling different tasks, it is often not necessary for all application models to be available. The use of model views serves not only for the simplified representation of complex issues but also to increase the security against unauthorized data access and is achieved by the simplest form of filtering through the use of multi-model templates.

\begin{figure}[htb]
\centering
\includegraphics[scale=0.5]{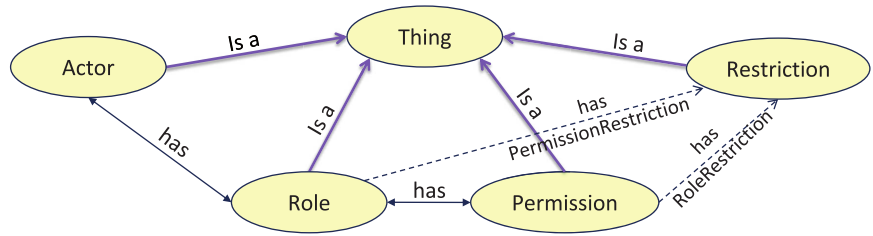}
\centering
\caption{filtering with multi-model templates}
\end{figure}

In the MEFISTO Project seven exemplary different multi-model templates have been defined which are based on the phases of the project and selected in the first step of each different subset of application models (Building, Operation, Performance, Costs and Construction Site Model). In a second step, the associated link model can be reduced to the remaining links of the tray model elements. Then, the templates demand not only the existence of application models in the multi-model, but claim additionally the compliance model-dependent levels of detail and processing status. It is conceivable that in a production model, the application building model (BM) and operation model (OM) are structured much more comprehensively than within the model range. In a third step, these additional conditions are checked using the metadata within the container. In Table 1, examples of granularities are outlined (e.g. in the application model the process model-column templates of the frame schedule (0.1), coarse schedule (0.2)) to detail schedule (1)). Other various project plans are discussed in more detail in [4].

\hspace{1 cm}
\begin{table}
\centering
\begin{tabular}
{|c|c|c|c|c|c|c|c|}
\hline
\textbf{Project Phase} & \textbf{Task} & \textbf{MM-Template} & \textbf{BM} & \textbf{OM} & \textbf{PM} & \textbf{CM} & \textbf{CSM} \\
\hline
Offer phase & Offer request & Tender Model & 0,5 & 0,1 & 0,1 & 0 & 0\\
\hline
Offer phase & Offer delivery & Offer  Model & 0,5 & 0,2 & 1 & 0,5 & 0\\
\hline
Contract  phase & Negotiation & Negotiation  Model & 0,5 & 0,2 & 1 & 0,7 & 0\\
\hline
Contract  phase & Commissioning & Contract  Model & 1 & 1 & 1 &  1  & 0\\
\hline
Execution  phase & Scheduling& Tender Model & 1 & 1 & 1 & 1 & 0,7\\
\hline

\end{tabular}
\caption{LOD of Application Models in Multi-model templates}
\end{table}

For the selection of individual object fields for further processing or to facilitate semantic multi-model filter, the use of specialized model cut-outs is desirable. Then these excerpts can turn into various, in part dynamic, criteria to be defined, for example as geometric part-models, time periods or (external) object properties of the model elements. The generation of task-specific ad-hoc multi-model views, which are combined to link models of BIM data with information from other application models is relatively complex and is a topic of current research. Katranuschkov describes in [7] an approach for the generation of BIM-based multi-model views, and a reference implementation, based on the use of open IFC toolset from the University of Weimar and HOCHTIEF that, with the use of the generic Model Subset Definition Schema (GMSD), focuses on these goals. Another challenge is the generation of partial geometric models, here under certain circumstances broken up into objects, because it must be generated at different levels of the object’s detail. This aspect is the subject of current research within the project HESMOS and is not described in this work.

\section{Context-dependent Access Control }
Camarinha-Matos differentiated in [8], exogenous and endogenous VO roles, with the former relationships reflecting outwards, such as interaction with the environment, customers, competitors and potential partners, and on the other side endogenous roles which are represented by the relationships within the VO. These endogenous roles handle the VO participants during the exercising of their cooperation and are defined in the VO-initiation, as well as other descriptive attributes, in the Organization Model. The classic role-based access control (RBAC) [9] defines the relationship in a simple way:

\hspace{1 cm}

\begin{center}
\textit{WHO (project role) may WHAT (authorizations)?}
\end{center}
\hspace{1 cm}

Hence, VO participants may bring different talents or skills into a VO, so that the VO actors are assigned with so-called potential roles. Both in [10] and in [11], it was found that the role assignment of the VO members is not rigidly fixed in practice, but contextual and dynamic role assignments are subject to restrictions and must therefore be context-dependently modelled and dynamically evaluated. Only with this combination, the observance of the principle of Separation of Duty [9] and the Principle of Least Privilege [12] can be ensured. We use an expanded role model for which the access control including the access context offers and determines illustrated in simplified form:

\hspace{1 cm}

\begin{center}
\textit{WHO (current role) may WHAT (authorizations) with WHOM (object) in which SITUATION (context)?}
\end{center}

\hspace{1 cm}

An example scenario: A player is entrusted with the tasks within the VO plan creation and plan evaluation. To ensure the Separation of Duty, this actor should not be allowed to control himself, so both roles can only be selected when accessing different objects.

\subsection{The use of context information }
To add context-sensitive abilities to security mechanisms, context attributes must be described in a formal context model. The context attributes are not individually described within this context model, but they are assigned real world objects. This so-called deputy approach describes a basic concept for the representation of contextual information [13] and link distributed context attributes of the subject and the object model together with the environment context attributes. Using a context model now we can combine previously individually held context information so that situations can be precisely described.
The relevant relationships for access decisions can be complex and involve various types and combinations of context information. For the modelling of the knowledge about their relationships, the impact of this information has to be formally described. For the definition of the access control policies, we use the Web Rule Language (SWRL) (Horrocks et al., 2003) to achieve flexibility and interoperability as well as easy administration. Some examples of SWRL rules:

\begin{itemize}[rightmargin=100 pt ,leftmargin=*,nolistsep]

\item  Role assignment:\\
(Actor(?a), Role(?e), Resource(?r), hasTarget(?a,?r), hasRole(?a,?e), hasOwner(?r,?a)) -\textgreater PermittedRole(?e)
\item        Ownership:\\
(Action(?a), Resource(?r), hasTarget(?a,?r), hasActor(?a,?x), hasOwner(?r,?x)) -\textgreater PermittedAction(?a)
\item  Sstate dependence:\\
(Action(?a), Resource(?r), hasTarget(?a,?r), hasObjectRestriction(?a,?x), hasPassRestriction(?r,?x)) -\textgreater PermittedAction(?a)
\end{itemize}

\subsection{Structure of the VO model}
We have modelled the RBAC-Model in our VO-Ontology model with the classes Actor, Role, Permission and Restriction. Objects of the Class Restriction define the context sensitive Assignment-Restriction in the form of SWRL-rules. The Relation sr is modelled by the bijective mapping of the ObjectPropertys hasActorRoleAssignment and hasRoleActorAssignment, which connect the classes Actor and Role.

\begin{figure}[htb]
\centering
\includegraphics[scale=0.60]{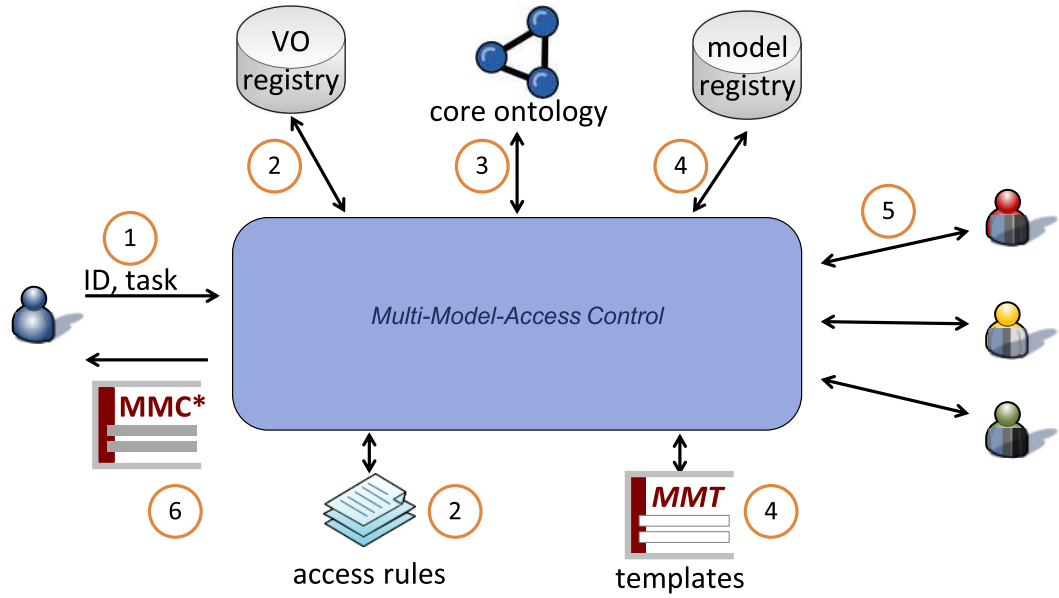}
\centering
\caption{VO Model}
\end{figure}

Both, the ObjectPropertys hasPermissionRoleAssignment and hasRolePermissionAssignment connect the classes Role and Permission and reflect the Relation sr. The SubPropertys hasRoleRestriction and hasPermissionRestriction link the Class Restriction to the ObjectPropertys hasActorRoleAssignment and hasPermissionRoleAssignment and realize the functions fc(sr) and gc(pr) (in Figure 3 dotted). The Restriction-Objects contain the restrictions to be evaluated, where the „Open World Assumption“ is valid, i.e. the modelled Relations sr and pr are true, if there are no Restrictions or the Restrictions can be evaluated to be true.

\subsection{Dynamic Authorization Decisions }
Scholz describes in [15] mutual trust as the basis for VOs and the transparent handling of explicit rules and standards as the basis for this confidence. For the rules of a context-based access control for multi-models, we use the Fine Grain Access Control (FGAC) rules described by Franzoni in [16] with a context-sensitive expansion, which additionally accesses the context in which access decisions are integrated. The actual access role when accessing an object instance from a dynamic access control according to the access context, consists of various determined static (potential roles with permissions) and dynamic conditions (object status and subject status, subject-object relationships and project status). This consists mainly of an empty MMC with pre-populated metadata descriptions. Using these templates it is easy to determine whether a padded MMC corresponds to a template. In MEFISTO a VO-based service platform is used which registers all created MMC templates and the metadata of all transmitted application models.

\hspace{1 cm}
\begin{figure}[htb]
\centering
\includegraphics[scale=0.75]{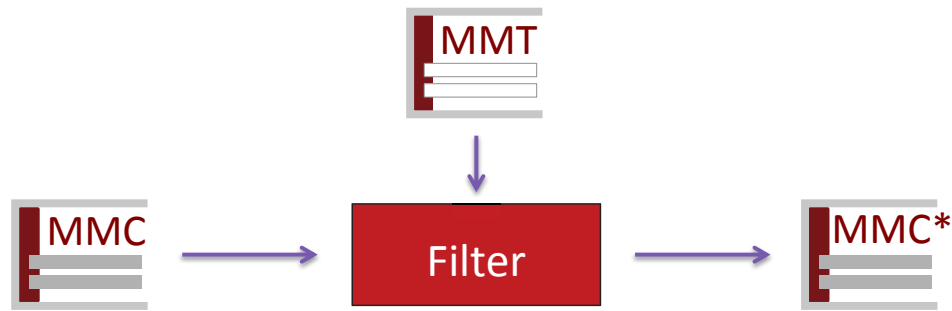}
\centering
\caption{Access Control with Multi-Model-Filterl}
\end{figure}

\hspace{1 cm}

We propose a context-sensitive access control in 5 steps:

\begin{enumerate}
\item The role of a requesting actor is determined by evaluating its potential VO roles and the access context.
\item The User ID is checked in the VO registry and connected access rules are evaluated.

\item  With the Task description in the core ontology, the necessary permissions are evaluated
\item  The task-dependent Template is searched on the VO-platform. If there is no template, then it will return an unfiltered Multi-Model, according to the preferences, or the access will not be permitted. If there is a template, it is created on the basis of a container template a Multi-Model-Container through the base service. Then, the registered trade models use the service from the platform and search in the model registry for an application model in the desired resolution.
\item  If an application model does not exist in the required granularity, the subject-specific model type of filter methods and services are requested and an application model with higher granularity and the desired level of detail are given as an argument. The application model produced is inserted into the Multi-Model-Container.
\item This Multi-Model-Container is transferred to the actor.
\end{enumerate}

\section{Related work}
The common standardised access model role-based access control model (RBAC) [9] is based on the concept of the user role as an intermediary between subjects and permissions and is purely subject-based. The properties of the objects have no effect on the access decision. Therefore, role assignments may be wrong after a state change of the entities (objects as well as subjects). Various proposals have extended the RBAC model with attribute-based access control capabilities [17] which also consider environment attributes (e.g. system status, time, date), subject attributes (e.g. age, location, proofs of identity) as well as object attributes (e.g. size, value, location, state of objects). Priebe et al. [18] proposes an attribute-based access control approach, but roles are subject- dependently modelled. Kumar et al. use in [19] context filters for role assignment, based on hash tables describing user context and object context-attributes. Franzoni et al. presents in [16] a fine-grained access control model, with focus on Semantic Databases, using ontologies to describe the subject-object relationships. Last but not least, Goslar describes a context model with dynamic real-world object linking in an economic focus [13]. Common to all these approaches is an attribute based assignment of subjects to roles while the roles-permissions assignment remains static.

\section{Outlook}
It has outlined an approach which covers the aspects mentioned in chapter 1:
(A1) Due to one filter of the Multi-Model-Container, the communication between the partners was simplified and accelerated. Example, it is now possible to use Multi-Model-Container also on mobile devices. (A2) Only those application models necessary for the current processing situation were transferred to the Multi-Model-Container. (A3) The transferred application models have the exact resolution (size) needed for further processing.

\parskip 12pt
Both the major components of the filtering services as well as most of the required multi-model templates are still under development, making the approach only roughly sketched. Currently in the focus of the research at the TU Dresden are the methods for the generation of geometric partial models and scheduling time spans.

\parskip 12pt
Another development is focused on the filtering of MMC for transitive object properties of model elements. An example scenario for this comes from subsequent processing. Here, the complete reduction of MMC to a supplementary multi-model is desired which contains only subsequent positions in the application model specifications, corresponding to a reduced geometry and link model. Such application model sections, due to the dynamic component, cannot be generated with template models.

\nocite{*}
\bibliographystyle{eptcs}
\bibliography{generic}
\end{document}